\documentclass[pdflatex,sn-mathphys-num,iicol]{sn-jnl}

\usepackage{graphicx}%
\usepackage{multirow}%
\usepackage{amsmath,amssymb,amsfonts}%
\usepackage{amsthm}%
\usepackage{mathrsfs}%
\usepackage[title]{appendix}%
\usepackage{xcolor}%
\usepackage{textcomp}%
\usepackage{manyfoot}%
\usepackage{booktabs}%
\usepackage{algorithm}%
\usepackage{algorithmicx}%
\usepackage{algpseudocode}%
\usepackage{listings}%
\usepackage{longtable}
\usepackage{tabularx}
\usepackage{bm}
\usepackage{hyperref}

\begin{document}
\begin{titlepage}
\title[Article Title]{LegalMALR:Multi-Agent Query Understanding and LLM-Based Reranking for Chinese Statute Retrieval}

\author[1,2]{\fnm{Yunhan} \sur{Li}}\email{D24092110205@cityu.edu.mo}

\author[3,4]{\fnm{Mingjie} \sur{Xie}}\email{suat25060313@stu.suat-sz.edu.cn}

\author[4]{\fnm{Gaoli} \sur{Kang}}\email{kanggaoli@suat-sz.edu.cn} 

\author[2,3]{\fnm{Zihan} \sur{Gong}}\email{zh.gong2@siat.ac.cn}

\author[1]{\fnm{Gengshen} \sur{Wu}}\email{gswu@cityu.edu.mo}

\author*[2]{\fnm{Min} \sur{Yang}}\email{min.yang@siat.ac.cn}

\affil[1]{\orgdiv{Faculty of Data Science}, \orgname{City University of Macau}, \orgaddress{\street{Avenida Padre Tomás Pereira Taipa}, \city{Macau}}}

\affil*[2]{\orgdiv{Shenzhen Key Laboratory for High Performance Data Mining, Shenzhen Institutes of Advanced Technology}, \orgname{Chinese Academy of Sciences}, \orgaddress{\street{Xueyuan Avenue}, \city{Shenzhen} \state{Guangdong Province}, \country{China}}}

\affil[3]{\orgname{Southern University of Science and Technology}, \orgaddress{\street{1088 Xueyuan Avenue, Nanshan District}, \city{Shenzhen}, \state{Guangdong}, \country{China}}}

\affil[4]{\orgdiv{Artificial Intelligence Research Institute}, \orgname{Shenzhen University of Advanced Technology}, \orgaddress{\street{No. 1 Gongchang Road, Guangming District}, \city{Shenzhen}, \state{Guangdong},\country{China}}}


\abstract{Statute retrieval is essential for legal assistance and judicial decision support, yet real-world legal queries are often implicit, multi-issue, and expressed in colloquial or underspecified forms. These characteristics make it difficult for conventional retrieval-augmented generation pipelines to recover the statutory elements required for accurate retrieval. Dense retrievers focus primarily on the literal surface form of the query, whereas lightweight rerankers lack the legal-reasoning capacity needed to assess statutory applicability.

We present LegalMALR, a retrieval framework that integrates a Multi-Agent Query Understanding System (MAS) with a zero-shot large-language-model-based reranking module (LLM Reranker). MAS generates diverse, legally grounded reformulations and conducts iterative dense retrieval to broaden candidate coverage. To stabilise the stochastic behaviour of LLM-generated rewrites, we optimise a unified MAS policy using Generalized Reinforcement Policy Optimization(GRPO). The accumulated candidate set is subsequently evaluated by the LLM Reranker, which performs natural-language legal reasoning to produce the final ranking.

We further construct CSAID, a dataset of 118 difficult Chinese legal queries annotated with multiple statutory labels, and evaluate LegalMALR on both CSAID and the public STARD benchmark. Experiments show that LegalMALR substantially outperforms strong Retrieval-augmented generation(RAG) baselines in both in-distribution and out-of-distribution settings, demonstrating the effectiveness of combining multi-perspective query interpretation, reinforcement-based policy optimisation, and large-model reranking for statute retrieval.
}


\maketitle
\end{titlepage}

\maketitle


\section{Introduction}\label{sec:intro}

Legal information retrieval is a central component of modern legal assistance, compliance auditing, and judicial decision support systems~\cite{li2025llmes}. Unlike general open-domain search, legal retrieval must operate over long and highly structured statutory texts and reason about complex relationships between factual conditions and statutory elements~\cite{maxwell2008concept, su2024stard}. In real-world settings, many user queries describe factual scenarios rather than citing statutory terminology directly, resulting in vague, narrative, or otherwise indirect expressions~\cite{su2024stard}. Because the surface form of a query often omits key legal elements and may implicitly reference multiple statutory issues, accurate retrieval requires going beyond keyword or embedding similarity and reconstructing the latent doctrinal structure of the query.

Existing retrieval-augmented generation (RAG) pipelines face several limitations in this context. First, dense retrieval modules typically operate on the original user query without considering that a single legal question may implicitly contain multiple legal issues or multiple statutory elements~\cite{vsavelka2022legal, locke2022case}. As a result, dense retrievers frequently fail to capture deeply embedded legal concepts, leading to suboptimal recall on complex queries. Second, single-model query reformulation often fails to recover missing factual or doctrinal conditions when the original query is implicit or underspecified~\cite{amato2025comparing}. Third, the reranking stage of standard RAG pipelines is limited by the reasoning capacity of lightweight cross-encoder rerankers~\cite{vsavelka2022legal}, which struggle to evaluate statutory applicability that depends on factual element satisfaction, conditional structures, and doctrinal coherence. These limitations collectively prevent traditional RAG systems from achieving high recall and accurate statutory relevance assessment in legal retrieval tasks.

\begin{figure}[t]
    \centering
    \includegraphics[width=0.40\textwidth]{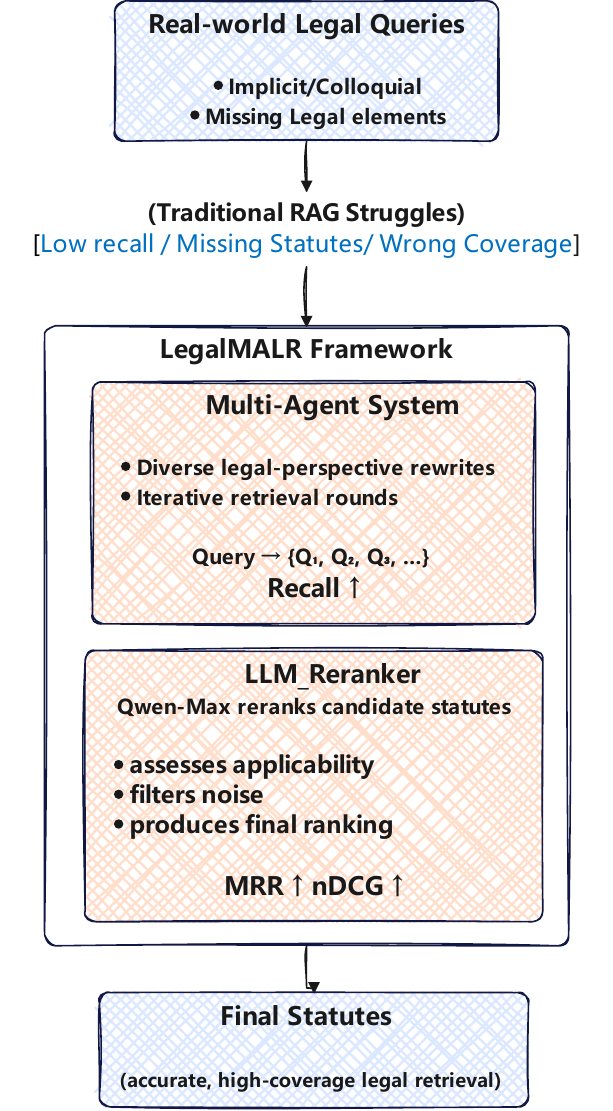}
    \caption{Overview of LegalMALR.
MAS expands retrieval coverage via diverse legal-perspective reformulations and iterative search, while the LLM eranker conducts statutory reasoning to produce accurate final rankings for implicit and colloquial legal queries.
    }
    \label{fig:mas_sample}
\end{figure}

Accurate statute retrieval therefore requires a system that exposes the implicit structure of a query and evaluates statutory applicability at a level deeper than semantic similarity. Empirical evidence from our ablation studies and case analyses, presented later in this paper (Sections~\ref{sec:experiments} and Appendix~\ref{app:case_study}), shows that reformulations generated by the same LLM under different agent roles and iterative trajectories tend to highlight different factual or doctrinal elements, and that some of these reformulations align far more closely with statutory requirements than others. This observation indicates that relying on a single reformulation is insufficient, and that exploring multiple complementary interpretations is essential for achieving high recall in complex legal queries. At the same time, the stochastic behaviour of LLM outputs introduces significant variability across reformulations, suggesting that optimisation techniques are needed to encourage more stable and legally grounded reformulation strategies.Finally, because statutory relevance depends on issue–element alignment rather than surface similarity, a more capable ranking mechanism is required at the reranking stage. Figure~\ref{fig:mas_sample} summarises these observations and motivates the design of our LegalMALR framework.

To operationalise these insights, we develop \textbf{LegalMALR}, a retrieval framework that integrates a \textbf{Multi-Agent Query Understanding System (MAS)} with a \textbf{zero-shot LLM-based reranking module (LLM Reranker)}. The MAS produces diverse reformulations of the query through a set of specialised agents that rewrite, decompose, and analyse the input from complementary legal perspectives. Because these agents are instantiated with Qwen-3-4B-Instruct, their outputs exhibit inherent variability. We therefore optimise the unified MAS policy using Generalized Reinforcement Policy Optimization (GRPO) to improve its stability and to encourage reformulations that better expose statutory applicability conditions. The reformulated queries trigger iterative dense retrieval calls whose merged candidate set is then evaluated by a commercial LLM, Qwen-Max, which serves as the \textbf{LLM Reranker} and performs natural-language legal reasoning to determine substantive relevance. Our study focuses on Chinese statutes and Chinese-language queries, and the datasets, retrieval components, and optimisation methods are designed for this statutory and linguistic context.

This work makes four main contributions. First, we introduce the CSAID dataset, a collection of 118 difficult Chinese legal queries annotated with multiple statutory labels, providing a realistic benchmark for evaluating retrieval systems under implicit and multi-issue legal reasoning settings. Second, we propose a multi-agent query understanding framework that generates diverse and complementary interpretations of a query and supports iterative retrieval. Third, we develop a GRPO-based optimisation method that improves the stability and retrieval effectiveness of the MAS policy. Fourth, we incorporate a Qwen-Max-based LLM Reranker that performs substantive legal reasoning to assess statutory relevance. Together, these contributions provide a coherent approach to the challenges posed by implicit, multi-faceted, and statute-level legal retrieval.

\section{Literature Review}\label{sec:liter_review}

Legal information retrieval covers tasks such as statute retrieval, legal question answering, and document relevance assessment. Prior work highlights that statutory texts differ significantly from open-domain corpora due to their hierarchical structure, dense cross-references, and reliance on domain-specific legal elements that determine applicability conditions~\cite{vsavelka2022legal, arslan2024survey}. Traditional lexical retrieval methods such as BM25~\cite{kim2022legal} and more recent dense retrieval models, including domain-adapted encoders like LegalBERT~\cite{chalkidis2020legal}, have improved semantic matching but still struggle when user queries lack explicit legal terminology or involve multiple statutory provisions. The STARD benchmark~\cite{su2024stard} further demonstrates that statute retrieval requires recognizing multi-element legal issues and aligning factual descriptions with statutory requirements, tasks that go beyond surface-level semantic similarity. These challenges motivate retrieval approaches that can better expose the latent structure of legal queries and support statute-level legal reasoning.

RAG has become a widely adopted paradigm for knowledge-intensive NLP, combining a dense retrieval module with a reranking component to provide high-quality evidence for downstream reasoning tasks~\cite{lewis2020retrieval, khattab2020colbert}. Dense retrievers such as DPR~\cite{karpukhin2020dense}, Contriever~\cite{izacard2022unsuperviseddenseinformationretrieval}, and ColBERT~\cite{khattab2020colbert} encode semantic similarity more effectively than lexical methods, while cross-encoder rerankers refine candidate results through token-level interactions. Recent large-model–based retrievers and rerankers, such as Qwen3-Embedding and Qwen3-Reranker~\cite{zhang2025qwen3}, have further advanced retrieval quality and become strong open-source baselines in general-domain RAG pipelines. In the legal domain, specialized models such as Infly’s inf-retriever-v1~\cite{inflyai2025inf} have also been introduced to address statute- and case-specific semantic patterns, achieving state-of-the-art performance on several legal retrieval benchmarks. RAG-style architectures have been applied to tasks such as legal question answering and decision prediction~\cite{chalkidis2024investigating, li2025llmes}, yet they exhibit two structural limitations when used for statute retrieval. First, the retrieval stage typically relies solely on the original user query, which is often insufficient when the query omits explicit legal terminology or embeds multiple latent issues that correspond to distinct statutory provisions. Second, conventional rerankers focus primarily on semantic similarity rather than legal element alignment, making them inadequate for evaluating statutory applicability in multi-element legal scenarios. These limitations indicate that statute retrieval requires stronger mechanisms for query understanding and more capable reranking strategies than those provided by standard RAG pipelines.

Query reformulation has long been used to improve retrieval effectiveness in information retrieval, with classical approaches such as pseudo-relevance feedback and query expansion aiming to enrich or disambiguate the user query~\cite{lavrenko2017relevance, carpineto2012survey}. LLMs have recently enabled more powerful reformulation strategies by generating semantically diverse interpretations of a query. Techniques such as multi-query rewriting, query decomposition, and step-back prompting improve retrieval coverage by surfacing complementary semantic perspectives that may not appear in the original input~\cite{shen2023large, zheng2023take, amato2025comparing}. These methods are particularly useful for legal retrieval, where queries often contain multiple factual elements or doctrinal requirements that must be separated or abstracted before the relevant statutes can be identified.

Beyond single-model rewriting, recent research has explored using multiple LLM agents to produce diverse reasoning trajectories or complementary reformulations~\cite{du2023improving, li2025agent4ranking}. Multi-agent collaboration helps surface different latent aspects of a query and reduce single-model biases, yet existing approaches typically rely on heuristic aggregation, and the quality of agent outputs varies substantially due to the inherent stochasticity of LLMs. To address this instability, several works in general-domain~\cite{chen2025improvingretrievalaugmentedgenerationmultiagent}setting and in medical-domain retrieval~\cite{xu2025rar} have begun applying reinforcement learning or preference optimization to multi-agent RAG systems, demonstrating notable gains in robustness and retrieval quality. However, according to our review of the literature, such optimization-oriented multi-agent methods remain rare in statute-level legal retrieval, where queries are highly implicit and statutory relevance depends on complex legal-element alignment. These gaps highlight the need for methods that can systematically explore, optimize, and stabilize multi-agent query understanding for legal retrieval.

Reranking is a crucial component of retrieval pipelines, refining candidate lists through deeper query–document interaction. Early neural rerankers, including cross-encoder models such as MonoBERT, established strong baselines by modeling token-level similarity~\cite{nogueira2019multistagedocumentrankingbert}, but they lack the multi-step reasoning capabilities required for complex decision-making tasks. With the emergence of LLMs, recent work has explored using LLMs as general-purpose rerankers capable of contextual comparison and justification-based relevance assessment. Among these, RankGPT~\cite{sun2024chatgptgoodsearchinvestigating} provides one of the closest methodological precedents to our work: it frames reranking as an instruction-following task and uses GPT-4-class models in a zero-shot, listwise setting, demonstrating that large LLMs can outperform traditional cross-encoders without task-specific fine-tuning. Other studies, such as RankFlow~\cite{jin2025rankflow}, further outline multi-role LLM-based reranking workflows, though they have not been validated in legal contexts. In parallel, high-performance open-source rerankers such as Qwen3-Reranker~\cite{zhang2025qwen3} fine-tune smaller 0.6B–8B models to approximate LLM-level ranking quality, but their reasoning capacity remains constrained by model size. Collectively, these developments reflect a field-wide transition from similarity-driven scoring toward reasoning-driven reranking architectures powered by large LLMs.

In legal retrieval, the limitations of conventional rerankers—including small LLM-based rerankers—become more pronounced. Statutory relevance requires aligning the factual elements present in a query with the applicability conditions defined within a statute, often involving multi-element legal tests, nested exceptions, and hierarchical clause structures that go beyond surface-level semantic matching~\cite{vsavelka2022legal, chalkidis2020legal}. Smaller cross-encoders or 4B-scale rerankers frequently struggle to model these implicit dependencies, particularly when statutory applicability hinges on subtle doctrinal distinctions or multi-condition reasoning. Large commercial LLMs, by contrast, possess significantly stronger capabilities for conditional evaluation and structured legal reasoning, yet their use as zero-shot statute-level rerankers remains limited in the existing literature. These gaps underscore the need for reranking mechanisms that explicitly leverage the reasoning ability of large LLMs to assess statutory applicability and support more accurate legal retrieval.

\section{Methodology}\label{sec:Methodology}

\begin{figure*}[t]
    \centering
    \includegraphics[width=0.95\textwidth]{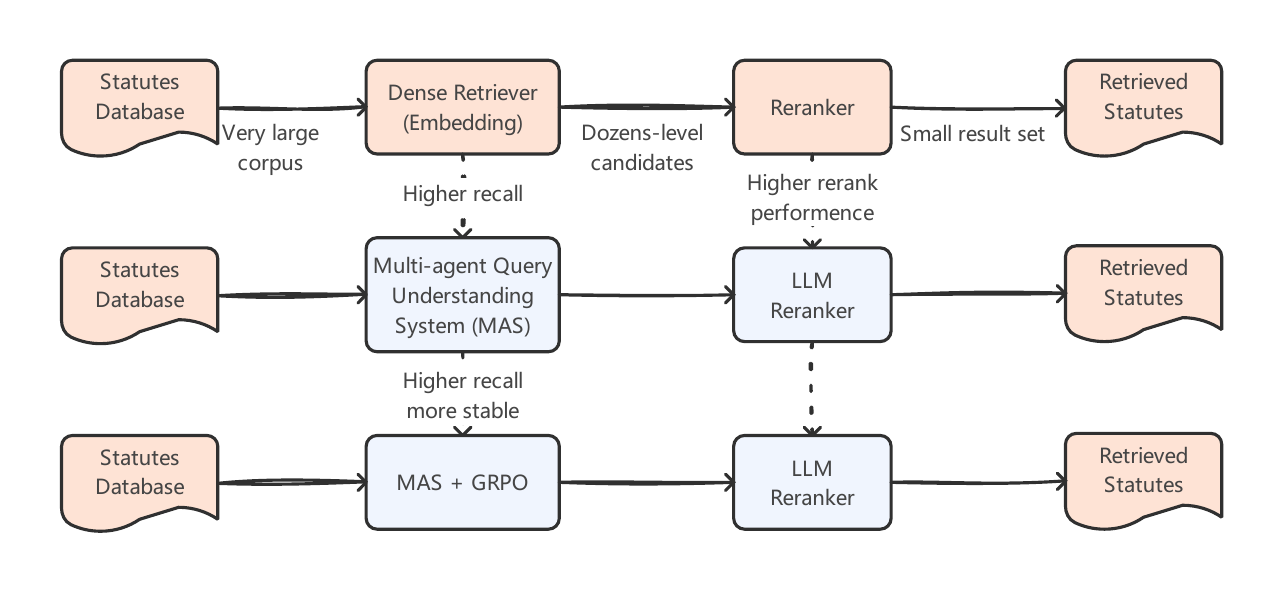}
    \caption{
    Overall architecture of LegalMALR. 
    The upper row illustrates the conventional dense-retrieval pipeline in which a single embedding retriever issues one-shot retrieval over the user query. 
    The middle row presents the proposed Multi-Agent Query Understanding System (MAS), which replaces this single retrieval decision with multi-perspective, multi-round reformulation and retrieval. 
    The lower row depicts the complete LegalMALR framework, where MAS is further stabilised through GRPO and the merged candidate set is reranked by a large language model.
    }
    \label{fig:architecture}
\end{figure*}

This section introduces the LegalMALR framework, designed to address the challenges posed by real-world legal queries that often contain implicit, multi-element, or colloquial structures. As shown in Figure~\ref{fig:architecture}, LegalMALR departs from standard dense retrieval by incorporating a multi-agent query understanding stage and a large-model-based reranking mechanism. The following subsections summarise the system at three levels: a conceptual overview, the detailed end-to-end processing pipeline, and the concrete model components used in each stage.

\subsection{Overview of LegalMALR}\label{sec:overview}

\begin{figure}[!htbp]
    \centering
    \includegraphics[width=0.45\textwidth]{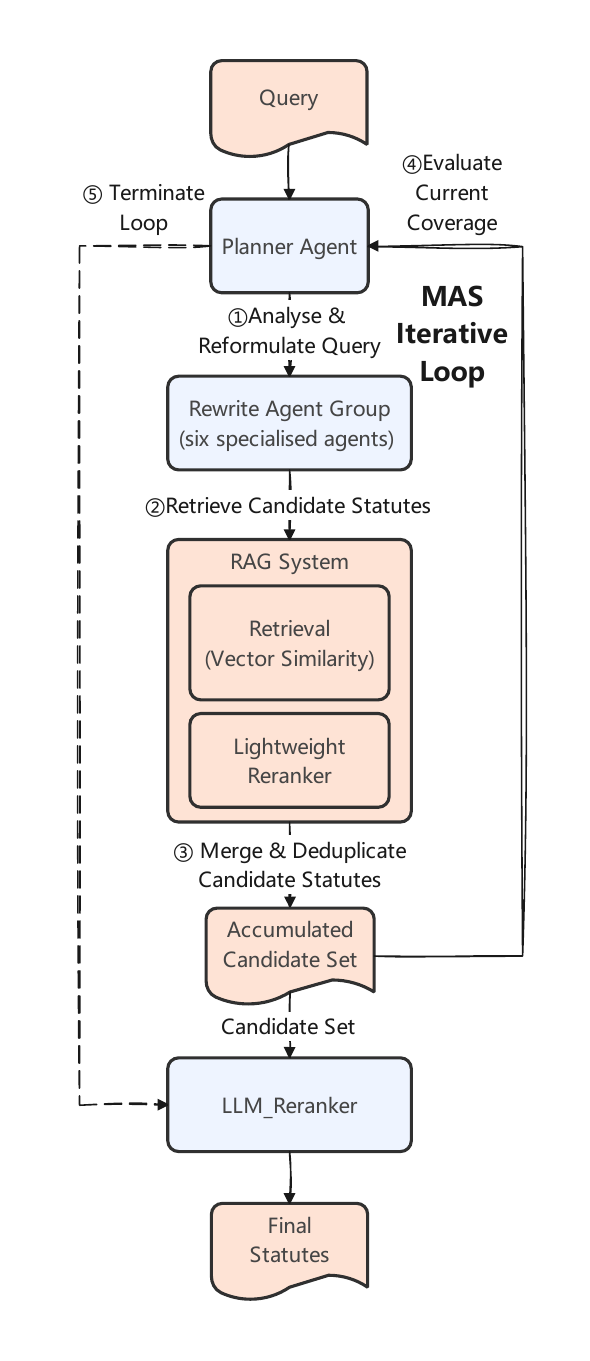}
    \caption{
    End-to-end processing flow of LegalMALR. 
    MAS iteratively analyses and reformulates the query, retrieves candidate statutes, and monitors coverage to decide termination. 
    The final merged candidate pool is evaluated by a zero-shot LLM-based reranker to produce the ranked statute list.
    }
    \label{fig:malr_pipeline}
\end{figure}

LegalMALR replaces the traditional single-query dense retrieval stage with a Multi-Agent Query Understanding System (MAS) that generates multiple complementary reformulations of the input query. Each reformulation highlights a different legally plausible interpretation of the underlying scenario, allowing the system to explore a broader portion of the statute space. These reformulations are individually processed by an embedding-based retriever and a lightweight neural reranker, thereby constructing a recall-oriented candidate pool that captures statutes semantically distant from the literal query. Because LLM-generated reformulations naturally exhibit behavioural variability, we further optimise the MAS policy using GRPO to encourage stable, structured, and legally meaningful reformulation trajectories. After MAS concludes, the accumulated candidate set is evaluated by a commercial LLM (Qwen-Max) acting as a zero-shot reranker, which assesses statutory applicability based on factual alignment, element satisfaction, exception structures, and doctrinal coherence. This division between exploratory recall (MAS) and precision-oriented legal judgement (LLM Reranker) enables LegalMALR to achieve high recall without sacrificing legal reasoning quality. The overall end-to-end workflow is illustrated in Figure \ref{fig:malr_pipeline}.

\subsubsection{End-to-End Processing Pipeline}\label{sec:pipeline}

Given a user query, MAS initiates an iterative cycle controlled by a Planner Agent. In each iteration, the Planner selects one specialised rewrite agent to generate a legally informed reformulation, which is then submitted to the retrieval module. Retrieval is performed through a dense vector encoder (Qwen3-Embedding-4B) followed by a lightweight neural reranker (Qwen3-Reranker-4B). Retrieved statutes are merged and deduplicated into an expanding candidate pool. After each iteration, the Planner evaluates the coverage and diversity of the current pool and determines whether further reformulations are required. This dynamic stopping rule enables MAS to perform only a single iteration for simple queries while allocating multiple rounds of exploration for more complex or ambiguous ones. Once MAS terminates, the final candidate set is passed to the zero-shot LLM Reranker (Qwen-Max), which conducts substantive statutory reasoning and outputs the ultimate ranked list.

\subsubsection{Model Components}\label{sec:model_components}

LegalMALR uses three groups of models corresponding to the three stages of the pipeline. (1) MAS Backbone. All planner and rewrite agents are instantiated from the same instruction-tuned model, \textbf{Qwen3-4B-Instruct-2507}, with GRPO applied to obtain a unified MAS policy. (2) Retrieval Components. Each reformulated query is processed by \textbf{Qwen3-Embedding-4B} (dense retriever) and \textbf{Qwen3-Reranker-4B} (lightweight reranker), both used without task-specific finetuning. (3) Final LLM Reranker. A commercial large model, \textbf{Qwen-Max}, serves as a zero-shot statutory relevance assessor accessed via API and used solely for inference-time reranking.

\subsection{Dataset}\label{sec:dataset}

This study evaluates statute retrieval under two Chinese-language datasets constructed within the Mainland China civil-law system. Although both datasets involve Chinese statutory interpretation, they differ markedly in realism, implicitness, and annotation scope. The first is the proposed \textbf{CSAID} (Chinese Statute Ambiguity and Implicitness Dataset), a collection of difficult real-world queries designed to capture implicit factual elements, domain-specific jargon, ambiguous phrasing, and multi-statute reasoning patterns. The second is the public \textbf{STARD} dataset~\cite{su2024stard}, which provides a controlled benchmark setting with systematically annotated statute-level relevance labels. The following subsections describe these datasets in detail, and representative CSAID examples are provided in Appendix~\ref{app:CSAID}.

\subsubsection{The CSAID}\label{sec:CSAID}

The CSAID is constructed to evaluate statute retrieval systems under realistic Chinese-language legal search scenarios. It contains 118 anonymized queries obtained from a commercial legal service environment and reflects the ambiguity, implicitness, and domain-specific conventions characteristic of practical legal inquiries in the Mainland China civil-law context. All personal identifiers in real-user queries were removed during anonymization. The dataset integrates three complementary sources:

\begin{itemize}
    \item \textbf{Real user queries.} Sixty queries are sampled from the statute-retrieval module of a commercial legal platform after removing malformed inputs and non-substantive system-testing queries.
    \item \textbf{Constructed queries based on classical cases.} Thirty queries are written by legal researchers drawing on well-known Chinese legal cases. These queries intentionally incorporate ambiguity, hidden factual conditions, and colloquial phrasing that mimic common user behavior.
    \item \textbf{Documented retrieval failures.} Twenty-eight queries originate from user-reported failure cases in an existing retrieval product, representing scenarios where conventional semantic retrieval is known to perform poorly.
\end{itemize}

All queries undergo a multi-stage quality-control pipeline. Initial statutory annotations are produced by legal researchers on an internal annotation platform and are subsequently reviewed by licensed attorneys to ensure doctrinal accuracy. The dataset thus captures realistic characteristics such as implicit legal elements, unstated conditional structures, multiple potentially relevant statutes, and non-technical natural-language formulations. Two illustrative examples highlighting these patterns are included in Appendix~\ref{app:CSAID}.

\subsubsection{STARD}\label{sec:stard}

The STARD dataset~\cite{su2024stard} is a publicly available Chinese statute-retrieval benchmark constructed to simulate real-world legal consultation scenarios. Each instance contains a natural-language query posed in layperson style and a set of expert-annotated statutory articles that directly address the underlying legal issue. The dataset is designed to evaluate whether retrieval systems can identify relevant legal provisions from a large statute corpus when confronted with queries containing colloquial expressions, incomplete factual descriptions, or implicit legal elements.

STARD provides an official split comprising 1,234 training queries and 309 test queries. In this work, the training split is used to optimise the MAS policy via GRPO, while all in-distribution evaluations are conducted on the 309-query test split. As a controlled and widely used public benchmark, STARD complements the CSAID dataset by providing a stable evaluation environment in which retrieval performance can be compared against a broad range of previously reported baselines.

\subsubsection{Task Formulation}\label{sec:task}

We formulate statute retrieval as a ranking problem defined over a fixed corpus of statutory articles. For an input natural-language query $q$, the objective is to produce a ranked list of articles $\{d_{1}, d_{2}, \ldots, d_{K}\}$ drawn from a statute set $\mathcal{A}$, where higher ranked items correspond to provisions that are legally applicable to the situation described by $q$. Gold annotations provide a subset $G(q) \subset \mathcal{A}$ that represents all statutes identified by legal experts as applicable. Retrieval performance is evaluated using standard rank-based metrics, including Recall@K, Mean Reciprocal Rank (MRR), normalised Discounted Cumulative Gain (nDCG), and Hit@K.

The proposed framework contains two components with distinct training regimes. The MAS (Multi-Agent Query Understanding System) is the only component that undergoes task-specific optimisation, trained with GRPO using the 1,234 training queries of the STARD dataset as the optimisation environment. The reward signal reflects MAS-stage recall over the merged candidate set, encouraging stable and legally structured reformulations. The LLM Reranker operates in a zero-shot setting without task-specific fine-tuning and evaluates only the candidate statutes produced by the MAS according to their legal relevance and doctrinal applicability.

Evaluation is conducted on two datasets. The 309-query STARD test split provides a controlled public benchmark for assessing in-distribution generalisation under layperson-style queries. The CSAID collection serves as a complementary real-world evaluation environment designed to stress-test implicit factual structures, domain-specific jargon, and ambiguous formulations that pose challenges for conventional retrieval pipelines. All core experiments, including ablation studies, are conducted on both datasets to ensure robustness and reliability of the conclusions.

\subsubsection{Dataset Statistics}\label{sec:dataset_stats}

Table~\ref{tab:dataset_statistics} summarises the key statistics of the two datasets used in this study. STARD contains 1,234 training queries and 309 test queries, with relatively short layperson-style questions and an average of 1.76 relevant statutes per query. The statute corpus associated with STARD comprises 55,348 articles.

CSAID, in contrast, consists of 118 highly implicit queries with an average length of 39.66 characters and an average of 7.16 relevant statutes per query. Its statute corpus contains 79,055 articles, representing broad coverage of Chinese statutory law. The substantially larger number of gold-relevant statutes per query indicates that relevant evidence is more widely dispersed across the corpus, making single-shot dense retrieval far less reliable and increasing the need for multi-perspective reformulation and iterative search. Combined with the prevalence of implicit factual structures and domain-specific terminology, these characteristics make CSAID a significantly more challenging evaluation setting for statute retrieval systems.

\begin{table}[h]
\centering
\small
\begin{tabular}{lcc}
\toprule
\textbf{Statistic} & \textbf{STARD} & \textbf{CSAID} \\
\midrule
Train queries & 1,234 & --- \\
Test queries & 309 & 118 \\
Avg. query len. & 27.31 & 39.66 \\
Avg. statute len. & 126.80 & 170.77 \\
Statute corpus size & 55,348 & 79,055 \\
Avg. relevant statutes & 1.76 & 7.16 \\
\bottomrule
\end{tabular}
\caption{Summary statistics of the datasets used in this study. Statute length refers to the character length of statutory articles, and "relevant statutes" denotes the number of gold-annotated applicable statutes per query.}
\label{tab:dataset_statistics}
\end{table}

\subsection{Multi-Agent Query Understanding System}\label{sec:mas}

The first stage of the proposed framework aims to reconstruct the implicit legal structure of a user query before any statute-level reasoning is performed. Legal queries issued by non-professional users frequently omit key factual conditions, employ domain-specific jargon, or express multi-element legal issues in incomplete or colloquial forms. Such characteristics make direct retrieval from statutory corpora difficult, particularly in civil-law systems where the applicability of a statute depends on the satisfaction of a structured set of constituent elements.

To address these challenges, we introduce a MAS, which replaces the conventional single-query retrieval paradigm with an iterative, reformulation-driven multi-query retrieval process. Rather than issuing a single dense retrieval call based on the surface form of the query, MAS employs a coordinated set of specialised language agents to produce semantically complementary reformulations of the query, each capturing a distinct doctrinal or factual perspective. For each reformulation, the system retrieves candidate statutes using an embedding-based dense retriever and a lightweight neural reranker. These candidates are merged and deduplicated across iterations, allowing the system to accumulate a progressively more comprehensive and recall-oriented candidate set.

MAS operates iteratively. The \textbf{Planner Agent} selects an interpretive strategy for the current query state, and the specialised \textbf{Rewrite Agents} generate legally informed reformulations. The retrieval module then identifies potentially relevant statutory provisions via vector similarity and lightweight neural reranking. The Planner Agent subsequently evaluates the coverage of the accumulated candidate set and decides whether further iterations are required. This dynamic loop enables MAS to allocate computation adaptively, performing only a single iteration for simple queries while expanding the search space through multiple reformulations for more complex or ambiguous cases.

All agents in MAS share the same \textbf{Qwen-3-4B-Instruct-2507} backbone, representing the GRPO-optimised unified MAS policy, with differentiation arising solely from agent-specific system prompts. The following subsections describe the motivation, agent design, reformulation mechanism, and aggregation process in detail.

\subsubsection{Motivation: Latent Structures and Legal Ambiguity in User Queries}\label{sec:mas_motivation}

A central challenge in statute retrieval arises from the mismatch between the surface linguistic form of user queries and the doctrinal structures that determine statutory applicability. Layperson legal questions frequently contain incomplete factual descriptions, implicit conditions, and non-professional terminology~\cite{su2024stard, askari2024answer}, and these characteristics are especially prominent in real-world Chinese consultation data. Empirical evidence from the CSAID collection illustrates this phenomenon. For instance, Example~A in Appendix~\ref{app:CSAID} appears, from its literal phrasing, to concern wildlife-protection or food-safety legislation, yet its underlying legal issue involves the defence of necessity in criminal law, which depends on unstated conditions such as imminent danger and proportionality. A different form of ambiguity is reflected in Example~B in Appendix~\ref{app:CSAID}, where an industry-specific expression implicitly encodes conditional obligations even though such meaning is absent from statutory language. Because the key legal implication is conveyed implicitly through jargon, conventional semantic retrieval struggles to surface the correct provisions.

These observations indicate that a single surface-level interpretation of a query is insufficient for reliable statute retrieval. Distinct plausible interpretations may correspond to different legal issues and statutory elements, and capturing these complementary perspectives typically cannot be achieved through a single rewrite or a single-pass dense retrieval operation. This motivates a multi-agent approach in which specialised agents reinterpret the query from complementary legal perspectives, generating diverse reformulations that expose latent conditions, clarify ambiguous terminology, and articulate the underlying legal relationships.

\subsubsection{Agent Roles and Legal Interpretation Perspectives}\label{sec:mas_roles}

The MAS operates through a collection of specialised agents, each designed to capture a distinct legal interpretation perspective during multi-round query processing. Working together, these agents dissect latent factual structure, clarify ambiguous expressions, and articulate alternative formulations that reflect different doctrinal viewpoints. These complementary perspectives enable exploration of a broad space of potentially relevant statutory interpretations and substantially increase retrieval coverage. All agents share the same \textbf{Qwen-3-4B-Instruct-2507} backbone, with differentiation arising solely from agent-specific system prompts; the functional specifications of these prompts are provided in Appendix~\ref{app:agent_prompts}. This design ensures computational comparability across agent choices, since all reformulations ultimately rely on the same underlying model capacity. The functional roles of all agents are summarised in Table~\ref{tab:mas_roles}.

\begin{table*}[t]
\centering
\caption{Functional roles of agents in the MAS system.}
\label{tab:mas_roles}
\begin{tabularx}{\textwidth}{p{3.0cm} X}
\toprule
\textbf{Agent} & \textbf{Primary Function} \\
\midrule
\textbf{Planner Agent} & Controls the MAS loop by analysing the query and intermediate retrieval results, selecting the next reformulation strategy, and determining termination. Each iteration activates exactly one Rewrite Agent. \\
\midrule
\textbf{Single-Element Rewrite Agent} & Clarifies colloquial or loosely structured expressions and rewrites them into precise legal terminology while preserving meaning. \\
\midrule
\textbf{Supplementary-Element Rewrite Agent} & Makes implicit but legally decisive factual or normative conditions explicit, including thresholds, actor identities, behavioural requirements, and contextual qualifiers. \\
\midrule
\textbf{Multi-Element Decomposition Agent} & Identifies structurally complex queries and decomposes them into multiple focused sub-queries, each independently triggering a retrieval call. Note that the Multi-Element Decomposition Agent may generate multiple reformulations within a single iteration, each leading to an independent retrieval call. \\
\midrule
\textbf{Supportive-Law Rewrite Agent} & Generates reformulations aimed at retrieving interpretive, procedural, or auxiliary provisions that support core substantive statutes. \\
\midrule
\textbf{Semantic-Abnormality Repair Agent} & Identifies semantic abnormalities such as domain overlap, latent principles, or procedural dependencies and produces a repaired, legally coherent reformulation. \\
\bottomrule
\end{tabularx}
\end{table*}

These agents collectively generate a diverse set of reformulations that support structural decomposition, factual clarification, doctrinal augmentation, and coherence repair. This diversity is essential for reconstructing the implicit legal issues embedded in real-world queries and for enabling the MAS to explore the statute space from multiple legally meaningful perspectives without performing full statutory reasoning.

\subsubsection{Reformulation Workflow and Candidate Accumulation}\label{sec:mas_workflow}

The MAS operates through an iterative reformulation-and-retrieval cycle that progressively expands the legal interpretation space of a query. At the beginning of each iteration, the Planner Agent evaluates the current coverage of the accumulated candidate pool and selects one specialised Rewrite Agent to act next. The selected agent then generates one or more reformulated queries depending on its functional role. Most agents produce a single reformulation per iteration, whereas the Multi-Element Decomposition Agent may output several focused sub-queries that each represent a distinct doctrinal facet of the underlying scenario.

Each reformulation produced during the iteration is independently submitted to the retrieval module, which performs dense vector-similarity search using an embedding-based retriever and then applies a lightweight neural reranking step to produce a compact list of candidate statutes. All retrieved candidates from the reformulations within the iteration are merged into a shared candidate pool through deduplication and normalisation. This aggregation mechanism allows MAS to accumulate statutes corresponding to multiple doctrinal possibilities, including those that are semantically distant from the surface form of the query but legally relevant under alternative interpretations.

Throughout the iterative process, the Planner Agent monitors the growth, diversity, and stability of the accumulated candidate set. Once additional reformulations are unlikely to yield substantive improvements in coverage, the Planner issues a termination decision. This coverage-based termination criterion results in dynamic retrieval depth, where simple queries often conclude after a single iteration while more complex or ambiguous queries trigger multiple rounds of reformulation and retrieval.

Upon termination, the MAS outputs a recall-oriented and legally enriched candidate set that serves as the input to the subsequent LLM Reranker stage. By decoupling exploratory recall from deep doctrinal reasoning, the MAS provides a stable and interpretable foundation for downstream statutory evaluation and enables robust performance on ambiguous, under-specified, and multi-faceted legal queries.

\subsection{LLM Reranker}\label{sec:llm_reranker}

The final stage of the retrieval pipeline applies a large language model based reranking module, denoted as the \textbf{LLM Reranker}. In this work, it is instantiated with the commercial \textbf{Qwen-Max} model, which operates in a zero-shot setting without any task-specific fine-tuning. This component receives the accumulated candidate statutes constructed by the MAS system and produces a ranked ordering according to their legal relevance. Whereas MAS is designed for broad, structurally diverse recall, the LLM Reranker focuses on fine-grained statutory applicability and directly influences ranking-oriented metrics such as MRR, nDCG, and Recall@K. The component is used strictly at inference time and is fully decoupled from GRPO optimisation.

Each candidate statute is provided to the reranker together with the original query. The model is instructed to evaluate doctrinal applicability, factual alignment, and conditional structures while outputting only a compact JSON object specifying the indices of the selected statutes. The prompt template used for this process is summarised in Appendix~\ref{app:reranker_prompt}. Importantly, although the instruction encourages deep legal assessment, the model is \emph{not} required to generate chain-of-thought rationales; only the final structured ranking is returned. This design maintains determinism, avoids hallucinated rationales, and ensures strict output-format stability essential for listwise reranking.

During development, we also experimented with enabling explicit chain-of-thought generation, in which the model first generates a short textual rationale before producing the final ranking. These preliminary tests indicated that explicit reasoning can yield modest improvements in ranking accuracy, particularly on complex multi-element queries. However, explicit rationale generation substantially increases inference cost (both latency and token usage), introduces stochastic variability in output formatting, and complicates fair comparison with baseline RAG pipelines. To preserve computational efficiency, output reliability, and evaluative consistency, all reported results employ the no-CoT configuration: the LLM conducts internal reasoning implicitly but outputs only the final ranking in JSON form.

Compared to lightweight supervised rerankers, which rely primarily on token-level similarity signals, the LLM Reranker leverages broader contextual understanding and doctrinal reasoning. Even without fine-tuning, it produces more consistent and legally coherent rankings across candidate statutes. As demonstrated in our experiments, the LLM Reranker consistently improves the performance of both dense-retrieval-based and MAS-based candidate pools, confirming its complementary role in pairing high-recall retrieval with principled legal relevance assessment.

As demonstrated in our experiments, the LLM Reranker consistently improves the performance of both dense-retrieval-based and MAS-based candidate pools, confirming its complementary role in pairing high-recall retrieval with principled legal relevance assessment. For clarity, all reported end-to-end results in this paper use the no-CoT configuration, and comparisons across methods are performed under this consistent inference setting.

\subsection{GRPO Optimisation for MAS Stability}\label{sec:grpo}

Generalized Reinforcement Policy Optimization (GRPO) is employed to improve the stability and effectiveness of the MAS system by directly optimising its sequential decision-making process. Each MAS trajectory—i.e., a full reformulation–retrieval sequence controlled by the Planner Agent—corresponds to a sequence of policy actions that ultimately affects the recall of the accumulated candidate set. Because the MAS exhibits inherent variability across rollouts due to the stochastic behaviour of large-language-model-driven actions, GRPO provides a principled mechanism for learning a policy that produces more reliable and legally structured reformulations. In our setting, the policy is optimised using the MAS-stage recall over the merged candidate set as the trajectory-level reward, consistent with the evaluation cutoff used in the MAS ablation studies.

\subsubsection{Motivation for Reinforcement-Based Optimisation}\label{sec:grpo_motivation}

To quantify the variability inherent in the MAS, we conduct eight independent rollouts for each query on a held-out subset of the STARD training split and compute per-query statistics over the resulting recall values. For every query, we record the maximum, mean, and minimum recall across its eight rollouts, and then average these statistics over all queries. As shown in Table~\ref{tab:mas_variance}, the average maximum recall reaches 0.8725, indicating that the MAS is capable of producing highly effective reformulations. However, the average mean recall is substantially lower at 0.8098, and the average minimum recall drops to 0.7511. This discrepancy between attainable upper-bound behaviour and realised average behaviour reflects the instability induced by stochastic LLM-driven actions.

\begin{table}[!htbp]
\centering
\caption{Per-query recall variability across eight independent MAS rollouts. For each query, the maximum, mean, and minimum recall over its eight rollouts are computed, and the table reports the averages of these statistics across all queries.}
\label{tab:mas_variance}
\begin{tabular}{lccc}
\toprule
{} & \textbf{Max} & \textbf{Mean} & \textbf{Min} \\
\midrule
MAS (8-rollout) & 0.8725 & 0.8098 & 0.7511 \\
\bottomrule
\end{tabular}
\end{table}

A finer-grained inspection further highlights this instability. Queries are grouped into behavioural categories based on characteristic recall patterns observed across the eight rollouts. As summarised in Table~\ref{tab:mas_categories}, categories two and three exhibit pronounced instability: queries in these groups either occasionally or never achieve full recall yet fluctuate considerably between runs. These cases represent precisely the type of sequential decision variability for which reinforcement-based optimisation is well suited. The MAS clearly possesses the capacity to discover high-quality reformulations for such queries but lacks a mechanism for consistently selecting them.

\begin{table}[!htbp]
\centering
\caption{Behavioural categories obtained from the recall distributions across eight MAS rollouts. The groups summarise different stability patterns observed in MAS behaviour.}
\label{tab:mas_categories}
\begin{tabular}{lcc}
\toprule
\textbf{Cat.} & \textbf{Description} & \textbf{Count} \\
\midrule
1 & Stable and always correct & 763 \\
2 & Occasionally correct but unstable & 186 \\
3 & Never fully correct but unstable & 78 \\
4 & Consistently incorrect & 207 \\
\bottomrule
\end{tabular}
\end{table}

These observations show that the MAS has substantial untapped potential. Its intrinsic upper-bound recall is significantly higher than its realised operating average, and a large portion of queries fall into behavioural regimes where sequential decision-making is unstable. GRPO is therefore introduced to learn a more stable reformulation policy that more reliably approaches the intrinsic performance ceiling of the MAS.

\subsubsection{GRPO Formulation and Reward Design}\label{sec:grpo_formulation}

We adopt a single unified policy $\pi_{\theta}$ to govern the entire MAS system. Rather than training separate models for the Planner Agent and the specialised Rewrite Agents, all MAS actions are generated by the same policy and are optimised jointly. Each execution of the MAS yields a trajectory $\tau_i = (s_{i,0}, a_{i,0}, s_{i,1}, a_{i,1}, \ldots, s_{i,T_i})$ containing at most $T_i \le 5$ planner-controlled iterations. This upper bound is a design constraint: the MAS permits up to four reformulation–retrieval iterations, after which a final exit iteration is required. Although the Multi-Element Decomposition Agent may trigger multiple retrieval operations within a single iteration, the number of planner iterations never exceeds four.

The state $s_{i,t}$ is represented implicitly in the prompt context and includes the original query, prior planner decisions, generated reformulations, the merged retrieval set accumulated so far, and the current iteration index. An action $a_{i,t}$ corresponds either to the Planner’s choice of reformulation strategy or to the reformulated query produced by the selected agent. When decomposition yields multiple reformulations within a single iteration, each reformulation is treated as an independent policy action associated with the same state $s_{i,t}$ and contributes its own log-probability term.

The reward function contains terminal and lightweight intermediate components. The terminal reward is the MAS-stage recall over the merged candidate set produced by trajectory $\tau_i$:
\begin{equation}
R_{\mathrm{final}}(\tau_i) = \mathrm{Recall}(\mathrm{Results}(\tau_i),\, G(q)).
\end{equation}
Each iteration incurs a step penalty
\begin{equation}
R_{\mathrm{step}}(t) = -0.05,
\end{equation}

discouraging unnecessary looping and keeping the number of dense-retrieval calls close to the empirical average of 2.01 observed on the evaluation datasets. Intermediate rewards incentivise useful expansions of relevant coverage:

\begin{equation}
R_{\mathrm{hit}}(t) = \alpha \cdot \Delta\mathrm{Cov}(t),
\end{equation}
where $\Delta\mathrm{Cov}(t)$ denotes the incremental increase in retrieved gold statutes relative to $G(q)$ and $\alpha$ is a small shaping coefficient. A fallback penalty of $-5$ is applied only when the MAS executes an invalid early termination—specifically, when the Planner issues an exit action before any retrieval has occurred. Errors arising from empty strings or malformed rewrites are handled separately by the JSON parsing layer and do not trigger this penalty.

The total trajectory-level reward is thus
\begin{equation}
R_i = R_{\mathrm{final}}(\tau_i)
    + \sum_{t=1}^{T_i} R_{\mathrm{hit}}(t)
    + \sum_{t=1}^{T_i} R_{\mathrm{step}}(t).
\end{equation}

For each query, GRPO samples $K = 8$ trajectories and applies group-wise normalisation:
\begin{equation}
\hat{R}_i = \frac{R_i - \mu_R}{\sigma_R + \epsilon}.
\end{equation}
The log-probability of a trajectory is
\begin{equation}
\log \pi_{\theta}(\tau_i)
= \sum_{t=1}^{T_i} \log \pi_{\theta}(a_{i,t} \mid s_{i,t}),
\end{equation}
with multiple reformulations within a single iteration contributing multiple action terms. GRPO maximises the group-normalised policy-gradient objective:
\begin{equation}\label{eq:grpo}
L_{\mathrm{GRPO}}(\theta)
= - \frac{1}{K} \sum_{i=1}^{K} \hat{R}_i \log \pi_{\theta}(\tau_i),
\end{equation}
a REINFORCE-style estimator that improves stability via group-wise normalisation and requires no value function or advantage estimator.

\subsubsection{Training Procedure}\label{sec:grpo_training}

GRPO training is performed by repeatedly sampling trajectory groups from the unified MAS policy. For each training query, the policy is sampled $K = 8$ times to generate a trajectory group, and training proceeds in mini-batches of $G = 2$ groups (i.e., 16 trajectories per update). Each trajectory contains at most four planner iterations plus a final exit iteration, and may include multiple reformulations within a single iteration when decomposition occurs. After computing trajectory-level rewards and applying group-wise normalisation, policy parameters are updated using the GRPO objective in Eq.~\eqref{eq:grpo}. Optimisation follows default AdamW settings with cosine decay over ten epochs.

During training, the backbone language model remains frozen, and only lightweight LoRA-based adaptation layers are updated through the GRPO objective. This parameter-efficient design yields stable optimisation and keeps computational cost modest. At inference time, the MAS executes a single deterministic rollout with reduced temperature, producing one trajectory per query without repeated sampling. All experiments are conducted on an eight-GPU RTX 4090 setup, and a full GRPO training run requires approximately sixteen hours.

\section{Experiments}\label{sec:experiments}

This section evaluates LegalMALR on both in-distribution and out-of-distribution statute retrieval settings. We first describe the experimental setup, including datasets, evaluation metrics, baselines, and implementation details. Results and ablation studies are presented in subsequent subsections.

\subsection{Experimental Setup}\label{sec:exper_setup}

\paragraph{Datasets.} We evaluate LegalMALR on two Chinese statute retrieval benchmarks. All training is performed exclusively on the STARD training split, and CSAID is used for evaluation only, ensuring a strict cross-domain test setting.

(1) \textbf{STARD} \cite{su2024stard}. STARD contains real-world layperson legal queries annotated with relevant statutes. Its 1234 training queries are used to train the MAS policy via GRPO, and its 309 test queries are used for in-distribution evaluation and comparison with previously reported retrieval baselines.

(2) \textbf{CSAID}. CSAID contains 118 challenging real-world or case-derived queries with implicit factual elements and domain-specific expressions. It shares the same retrieval formulation as STARD but differs substantially in linguistic characteristics and relevance distributions, making it suitable for assessing cross-domain generalisation.

\paragraph{Evaluation Metrics.} We report Recall@K, Mean Reciprocal Rank (MRR@K), normalised Discounted Cumulative Gain (nDCG@K), and HitRate@K. Let $d_{1},\ldots,d_{K}$ denote the returned ranked list for a query and $G$ the set of gold statutes.

\[
\mathrm{Recall}@K = \frac{|G \cap \{d_{1},\ldots,d_{K}\}|}{|G|}.
\]

\[
\mathrm{MRR}@K = \frac{1}{|Q|}\sum_{q\in Q}\frac{1}{\mathrm{rank}_{q}}
\]

\[
\mathrm{rank}_{q} = \min\{i \le K \mid d_{i}\in G_{q}\},
\]
with $\mathrm{rank}_{q}=K+1$ if no relevant statute appears within top K.

\[
\mathrm{nDCG}@K = \frac{\mathrm{DCG}@K}{\mathrm{IDCG}@K}
\]

\[
\mathrm{DCG}@K = \sum_{i=1}^{K}\frac{\mathbb{I}(d_{i}\in G)}{\log_{2}(i+1)}.
\]

\[
\mathrm{HitRate}@K = \mathbb{I}\bigl(G \cap \{d_{1},\ldots,d_{K}\} \neq \varnothing\bigr).
\]

Unless otherwise specified, $K=10$ for end-to-end evaluation. Different experimental settings may adopt different values of $K$ and therefore the metric definitions remain generic.

For STARD, we follow the official protocol and report Recall@10 and MRR@10 because only these metrics are available in the public benchmark. For CSAID, which imposes no fixed protocol, we report Recall@10, MRR@10, nDCG@10, and HitRate@10.

\paragraph{Baselines.} We compare LegalMALR with five categories of retrieval baselines reported in the STARD benchmark: lexical matching (QL, BM25), open-domain PLMs (RoBERTa, SEED, coCondenser), legal-domain PLMs (SAILER, Lawformer), and fine-tuned dense retrievers (Dense-STARD, Dense-GPT4, Dense-CAIL, LSI-STARD). All results for these baselines are taken directly from the STARD publication.

In addition, we include three RAG pipelines built on the Qwen3 family: (1) \textbf{Qwen3-Original}, which combines Qwen3-4B-Embedding with Qwen3-4B-Reranker; (2) \textbf{Infly-Hybrid}, which combines the Infly inf-retriever-v1 dense encoder with Qwen3-4B-Reranker; and (3) \textbf{Qwen3-SFT}, a domain-supervised RAG pipeline obtained by finetuning the Qwen3-4B embedding and reranking models on legal-domain corpora. These models are selected because Qwen3-Embedding achieves state-of-the-art performance on multilingual embedding benchmarks such as MTEB and C-MTEB, and inf-retriever-v1 is the strongest publicly available legal-domain dense retriever on the Legal MTEB leaderboard as of 2025. Together these baselines represent general-purpose dense retrieval, legal-domain dense retrieval, and domain-supervised RAG pipelines.

To ensure fair comparison, all RAG systems including LegalMALR retrieve top 60 embedding candidates and rerank them into a final top 10 list.

\paragraph{Implementation Details.} LegalMALR consists of the MAS module and a final LLM-based reranker. The MAS policy is instantiated from Qwen3-4B-Instruct and trained with GRPO on eight NVIDIA RTX 4090 GPUs for approximately sixteen hours. The backbone model remains frozen during training and only lightweight LoRA-style adaptation layers are updated.

During inference, MAS executes a single rollout with fixed temperatures (0.6 for the Planner and 0.8 for rewrite agents), ensuring deterministic behaviour aside from controlled sampling induced by temperature settings. Each MAS iteration retrieves the top 30 embedding candidates and prunes them to the top 10 using the lightweight reranker. After multiple iterations, the merged and deduplicated candidate set contains on average 13.58 statutes, which are then passed to the final reranker.

\textbf{Dynamic Retrieval Budget and Fairness.} MAS uses an adaptive retrieval depth that terminates based on coverage signals. In practice, MAS performs between one and four retrieval rounds per query. Each round retrieves 30 dense candidates and retains 10 after lightweight reranking, resulting in 30 to 120 embedding calls. After merging, unique candidates range from 10 to 22 with an average of 13.58.

To ensure budget-equivalent comparison, we define retrieval cost by the number of dense embedding calls. Baseline RAG systems issue a single retrieval returning 60 candidates. Although MAS has a higher theoretical upper bound, its empirical average cost is 60.3 embedding calls, which is effectively identical to the fixed 60-call baseline. MAS therefore differs only in allocating this budget adaptively: simple queries stop early while complex queries invoke additional reformulations.

The final reranker is a commercial zero-shot model (Qwen-Max) that evaluates the MAS candidate set and outputs the final top 10 ranked statutes. The MAS candidate size is used only for MAS-stage ablation experiments and does not affect end-to-end evaluation.

\subsection{Experimental Results}\label{sec:main_results}

\begin{table*}[htbp]
\centering
\caption{Retrieval performance on the STARD test set (309 queries). Results for lexical, PLM-based, and fine-tuned dense retrievers are reproduced from the STARD benchmark~\cite{su2024stard}. All RAG baselines and LegalMALR use a matched retrieval budget of top--60 embedding candidates reranked to a final top--10 list.}
\label{tab:STARD_results}
\begin{tabular}{llcc}
\toprule
\textbf{Category} & \textbf{Method} & \textbf{Recall@10} & \textbf{MRR@10} \\
\midrule
\multirow{2}{*}{\textbf{Lexical Matching}} 
& QL~\cite{ponte2017language} & 0.4020 & 0.3304\\
& BM25~\cite{robertson2009probabilistic} & 0.3943 & 0.3369\\
\midrule
\multirow{3}{*}{\textbf{Open-Domain PLM}} 
& RoBERTa~\cite{cui2021pre} & 0.3908 & 0.3010\\
& SEED~\cite{lu2021less} & 0.3555 & 0.2816\\
& coCondenser~\cite{gao2022unsupervised} & 0.1598 & 0.1004\\
\midrule
\multirow{2}{*}{\textbf{Legal PLM}}
& SAILER~\cite{li2023sailer} & 0.3050 & 0.2234\\
& Lawformer~\cite{xiao2021lawformer} & 0.2989 & 0.2412\\
\midrule
\multirow{4}{*}{\textbf{Fine-tuned PLM}}
& Dense-STARD & 0.6061 & 0.4724\\
& Dense-GPT4 & 0.5174 & 0.4106\\
& Dense-CAIL & 0.1272 & 0.0842\\
& LSI-STARD & 0.2069 & 0.2156\\
\midrule
\multirow{3}{*}{\textbf{RAG system}}
& Qwen3-Original & 0.7579 & 0.6736\\
& Infly-Hybrid & 0.7233 & 0.6306\\
& Qwen3-SFT & 0.7690 & 0.7043\\
\midrule
\multirow{2}{*}{\textbf{Our}}
& LegalMALR & \textbf{0.8195} & \textbf{0.7367}\\
& $\Delta$ Qwen3-Original & $(+6.16)$ & $(+6.31)$\\
\bottomrule
\end{tabular}
\vspace{0.5ex}
\begin{minipage}{0.97\textwidth}
    \small
    \textit{Notes:} Non-RAG results follow the original STARD benchmark. All RAG systems and LegalMALR are evaluated under identical retrieval and reranking budgets.
\end{minipage}
\end{table*}

\begin{table*}[htbp]
\centering
\caption{Retrieval performance on the CSAID test set.}
\label{tab:CSAID_results}
\begin{tabular}{lcccc}
\toprule
\textbf{Method} & \textbf{MRR@10} & \textbf{Recall@10}& \textbf{nDCG@10}& \textbf{HitRate@10} \\
\midrule
Qwen3-Original & 0.8720 & 0.6032 & 0.6255 & 0.9661\\
Infly-Hybrid & 0.8171 & 0.5532 & 0.5679 & 0.8867\\
Qwen3-SFT & 0.8663 & 0.6323 & 0.6411 & 0.9746\\
\midrule
LegalMALR & \textbf{0.9161} & \textbf{0.6841}& \textbf{0.7126}& \textbf{0.9915}\\
$\Delta$ Qwen3-Original & $(+4.41)$ & $(+8.09)$ & $(+8.71)$ & $(+2.54)$\\
\bottomrule
\end{tabular}
\end{table*}

We evaluate LegalMALR under two complementary settings. The first setting examines in-distribution performance on the STARD test split, where the MAS policy is trained using the corresponding STARD training data. The second setting assesses cross-domain generalisation on the CSAID dataset, whose queries differ considerably from STARD in linguistic form, domain distribution, and the average number of relevant statutes. Both datasets share an identical retrieval formulation, enabling controlled comparison under distribution shift.

For STARD, we compare LegalMALR with all baseline categories introduced in Section \ref{sec:exper_setup}, including lexical matching methods, open-domain and legal-domain PLMs, fine-tuned dense retrievers, and three RAG pipelines. For CSAID, we report results for the three RAG baselines and LegalMALR. Non-RAG baselines reported in the STARD benchmark are designed for the STARD statute corpus and are not directly comparable when transferred to the CSAID environment. We therefore focus on RAG systems in the cross-domain evaluation. For all comparisons, Qwen3-Original is used as the primary reference system for reporting relative improvements.

\subsubsection{In-Distribution Results on STARD}

Table \ref{tab:STARD_results} presents retrieval performance on the 309-query STARD test set. Non-RAG baselines exhibit a substantial performance gap relative to modern neural retrieval pipelines. Lexical methods such as QL and BM25 achieve Recall@10 near 0.40. Open-domain PLMs including RoBERTa and SEED perform similarly or slightly worse, and coCondenser yields considerably lower performance. Legal-domain PLMs such as SAILER and Lawformer provide modest gains yet remain below 0.31 Recall@10. Among fine-tuned dense retrievers, Dense-STARD achieves the strongest performance, although the large variance across dense models highlights the sensitivity of these approaches to domain alignment and training data selection.

RAG baselines outperform all non-RAG systems. Qwen3-Original achieves 0.7579 Recall@10 and 0.6736 MRR@10, illustrating the strength of dense retrieval paired with a lightweight reranker. Infly-Hybrid performs slightly lower on STARD, suggesting that hybrid sparse dense configurations do not provide additional benefits in this benchmark setting. Qwen3-SFT, which incorporates supervised finetuning on legal-domain corpora, achieves 0.7690 Recall@10 and 0.7043 MRR@10 and serves as the strongest RAG baseline.

LegalMALR attains the best overall performance, achieving 0.8195 Recall@10 and 0.7367 MRR@10. These correspond to absolute gains of 6.16 percentage points in Recall@10 and 6.31 percentage points in MRR@10 over Qwen3-Original. LegalMALR also surpasses the domain-tuned Qwen3-SFT by a clear margin. These results demonstrate that the GRPO-optimised MAS policy produces more informative reformulations and higher-quality candidate sets, which directly improve the final LLM reranking outcome.

\subsubsection{Out-of-Distribution Results on CSAID}

Table \ref{tab:CSAID_results} reports results on the CSAID evaluation set, which represents a substantial distribution shift relative to STARD. CSAID queries contain more implicit factual elements, exhibit greater linguistic diversity, and on average correspond to a larger number of relevant statutes. These characteristics pose significant challenges for systems relying solely on dense retrieval.

Among the baseline RAG systems, Qwen3-Original achieves 0.8720 MRR@10 and 0.6032 Recall@10, indicating strong robustness despite the distribution shift. Infly-Hybrid performs consistently lower, suggesting reduced effectiveness of hybrid sparse dense retrieval when confronted with implicit or case-driven queries. Qwen3-SFT improves both ranking and recall accuracy, reaching 0.8663 MRR@10 and 0.6323 Recall@10, and serves as the strongest baseline under the CSAID setting.

LegalMALR achieves the highest performance across all metrics, obtaining 0.9161 MRR@10, 0.6841 Recall@10, 0.7126 nDCG@10, and 0.9915 HitRate@10. These represent absolute gains of 4.41, 8.09, 8.71, and 2.54 percentage points over Qwen3-Original. The improvements over Qwen3-SFT indicate that domain-specific finetuning alone is insufficient to capture the implicit and multi-element structures present in CSAID queries. By contrast, LegalMALR benefits from its GRPO-trained MAS policy, which dynamically produces diverse reformulations and yields a more complete candidate pool. This enables substantially improved generalisation under out-of-distribution conditions.

\subsection{Ablation: MAS and GRPO Training}\label{sec:mas_ablation}

\begin{table*}[t]
\centering
\caption{
Ablation study of the MAS module under zero-shot initialization and after GRPO training on the STARD and CSAID test sets. Results are evaluated using Recall@14 and HitRate@14, where $K=14$ corresponds to the average size of the MAS candidate pool after merging and deduplication. Bold values indicate the best performance for each dataset.
}
\label{tab:mas_grpo_ablation}
\begin{tabular}{lcccc}
\toprule
\multirow{2}{*}{\textbf{Model}} 
& \multicolumn{2}{c}{\textbf{STARD}} 
& \multicolumn{2}{c}{\textbf{CSAID}} \\
\cmidrule(lr){2-3} \cmidrule(lr){4-5}
& Recall@14 & HitRate@14 & Recall@14 & HitRate@14 \\
\midrule
Qwen3-Original  & 0.7845 & 0.8899 & 0.6319 & 0.9661 \\
Infly-Hybrid    & 0.7575 & 0.8576 & 0.6126 & 0.9745 \\
Qwen3-SFT       & 0.8129 & 0.9093 & 0.7090 & 0.9745 \\
MAS (zero-shot) & 0.8012 & 0.9094 & 0.6969 & 0.9831 \\
MAS + GRPO      & \textbf{0.8396} & \textbf{0.9320} & \textbf{0.7478} & \textbf{1.0000} \\
\bottomrule
\end{tabular}
\end{table*}

Table \ref{tab:mas_grpo_ablation} presents an ablation study evaluating the contributions of the MAS structure and the GRPO-optimised MAS policy. Unlike the end-to-end evaluation in Section \ref{sec:main_results}, which measures performance after the final LLM reranking, this ablation examines the quality of the candidate statutes produced directly by MAS. As described in Section \ref{sec:exper_setup}, MAS triggers an average of 2.01 dense-retrieval calls per query, computed over all evaluation trajectories on the STARD and CSAID datasets. Each retrieval call returns the top 30 embedding candidates and prunes them to the top 10 using the lightweight reranker. After merging and deduplication, the resulting candidate pool contains an average of 13.58 statutes. We therefore evaluate MAS-stage retrieval at $K = 14$. This metric isolates MAS performance before any contribution from the LLM reranker. MAS-stage recall is evaluated at the native candidate pool size (K=14), whereas all end-to-end metrics use K=10.

For reference, the table includes the three RAG baselines introduced earlier. Since their dense retrievers also output an initial candidate list prior to reranking, their Recall@14 and HitRate@14 provide a meaningful external comparison for evaluating MAS-stage recall quality under a budget-equivalent number of embedding calls.

Across both datasets, the zero-shot MAS already performs strongly, achieving Recall@14 values of 0.8012 on STARD and 0.6969 on CSAID. This places it close to Qwen3-SFT, a domain-supervised RAG baseline. These results indicate that the MAS structure by itself, through diverse reformulations and iterative exploration of the retrieval space, can surface relevant statutes that are difficult to obtain using a single-query dense retriever. Importantly, this behaviour is observed under a retrieval budget that is matched to the baselines in expectation, confirming that the improvements are not attributable to additional embedding calls.

Training MAS with GRPO yields consistent and substantial gains. MAS with GRPO achieves Recall@14 of 0.8396 and HitRate@14 of 0.9320 on STARD, outperforming both the zero-shot MAS and all RAG baselines at this retrieval stage. On CSAID, which features greater implicitness and broader semantic variability, MAS with GRPO further improves recall to 0.7478 and reaches a perfect HitRate@14 of 1.0. These results are aligned with the intended effect of GRPO, which encourages the MAS policy to generate reformulations that extend recall coverage and stabilise the retrieval process.

The contrast between zero-shot MAS and GRPO-trained MAS demonstrates that the MAS framework provides the structural flexibility required for multi-step retrieval, while GRPO supplies the optimisation signal that enables MAS to use this structure effectively. The more complete and higher-quality candidate sets produced by MAS with GRPO directly explain the improvements observed in the end-to-end evaluation in Section \ref{sec:main_results}. Overall, this ablation confirms that MAS and GRPO serve complementary roles: MAS expands the legal interpretation space through structured multi-agent reformulations, and GRPO refines this process into a reliable and distribution-aware retrieval policy.

\subsection{Ablation: Effect of the LLM Reranker}

\begin{table*}[t]
\centering
\caption{
Ablation of the LLM reranker under two independent retrieval settings: 
(1) a dense retriever producing sixty candidates (Qwen-E 60c), and
(2) the MAS with GRPO policy producing fourteen candidates (MAS 14c).
Within each retriever, the candidate set is fixed and only the reranker varies.
All results are reported with MRR@10, nDCG@10, and R@10, where R@10 denotes Recall@10.
Abbreviations: Qwen-E = Qwen3-4B-Embedding, Qwen-R = Qwen3-4B-Reranker,
LLM-R = zero-shot LLM-based reranker. Bold values indicate the best reranking
performance within each retriever setting.
}
\label{tab:llm_reranker_ablation}
\begin{tabular}{l l ccc ccc}
\toprule
\multirow{2}{*}{\textbf{Retriever}} 
& \multirow{2}{*}{\textbf{Reranker}} 
& \multicolumn{3}{c}{\textbf{STARD}} 
& \multicolumn{3}{c}{\textbf{CSAID}} \\
\cmidrule(lr){3-5} \cmidrule(lr){6-8}
& & MRR@10 & nDCG@10 & R@10 & MRR@10 & nDCG@10 & R@10 \\
\midrule
\multirow{2}{*}{Qwen-E 60c} 
& Qwen-R & 0.6305 & 0.6078 & 0.7230 & 0.8171 & 0.5678 & 0.5531 \\
& LLM-R  & \textbf{0.7546} & \textbf{0.7179} & \textbf{0.8101} & 
            \textbf{0.9248} & \textbf{0.6790} & \textbf{0.6329} \\
\midrule
\multirow{2}{*}{MAS 14c} 
& Qwen-R & 0.6853 & 0.6583 & 0.7763 & 0.8847 & 0.6539 & 0.6316 \\
& LLM-R  & \textbf{0.7367} & \textbf{0.7088} & \textbf{0.8195} & 
            \textbf{0.9161} & \textbf{0.7126} & \textbf{0.6841} \\
\bottomrule
\end{tabular}
\end{table*}

Table \ref{tab:llm_reranker_ablation} evaluates the contribution of the LLM-based reranker under two distinct retrieval conditions. The first condition uses a standard dense retriever based on Qwen3-4B-Embedding, which naturally produces a candidate list of sixty statutes. The second condition uses the MAS with GRPO policy, whose iterative retrieval process yields a more compact candidate set containing an average of fourteen statutes. These two settings constitute independent experimental conditions that reflect the characteristic behaviour of their respective retrievers. Within each condition, the candidate set is fixed and we substitute only the reranker. This two-by-two structure enables a controlled comparison of reranking quality without conflating retrieval-stage differences. Cross-condition results are therefore not interpreted as comparisons between the two retrievers.

Under the dense retrieval condition, replacing the lightweight Qwen3-4B-Reranker with the LLM-based reranker yields substantial improvements across all metrics on both datasets. On STARD, MRR increases from 0.6305 to 0.7546 and Recall@10 improves from 0.7230 to 0.8101. CSAID shows a similar pattern, with MRR improving from 0.8171 to 0.9248 and Recall@10 from 0.5531 to 0.6329. These gains indicate that the LLM reranker provides stronger semantic discrimination and is more capable of correcting noisy ordering in the dense retriever’s candidate list. Since the LLM reranker operates in a zero-shot setting without domain-specific finetuning, the improvements arise from its general reasoning ability rather than from additional supervision.

A consistent trend appears under the MAS with GRPO retrieval condition. When applied to the compact and higher-quality candidate lists produced by MAS, the LLM reranker continues to deliver clear improvements on both datasets. On STARD, MRR rises from 0.6853 to 0.7367 and Recall@10 from 0.7763 to 0.8195. On CSAID, MRR increases from 0.8847 to 0.9161 and Recall@10 from 0.6316 to 0.6841. These results show that the benefit of the LLM reranker is orthogonal to the retrieval strategy. Even when the candidate lists contain fewer irrelevant items, the LLM reranker is able to exploit finer semantic distinctions and extract additional performance gains.

Taken together, the two independent retrieval conditions demonstrate that the LLM reranker consistently strengthens both dense and multi-step retrieval pipelines. In conjunction with the MAS and GRPO results in Section \ref{sec:mas_ablation}, these findings show that LegalMALR’s end-to-end performance gains arise from complementary improvements in both retrieval and reranking.

\subsection{Ablation: One-step vs Multi-step Retrieval}

\begin{figure}[t]
    \centering
    \includegraphics[width=0.45\textwidth]{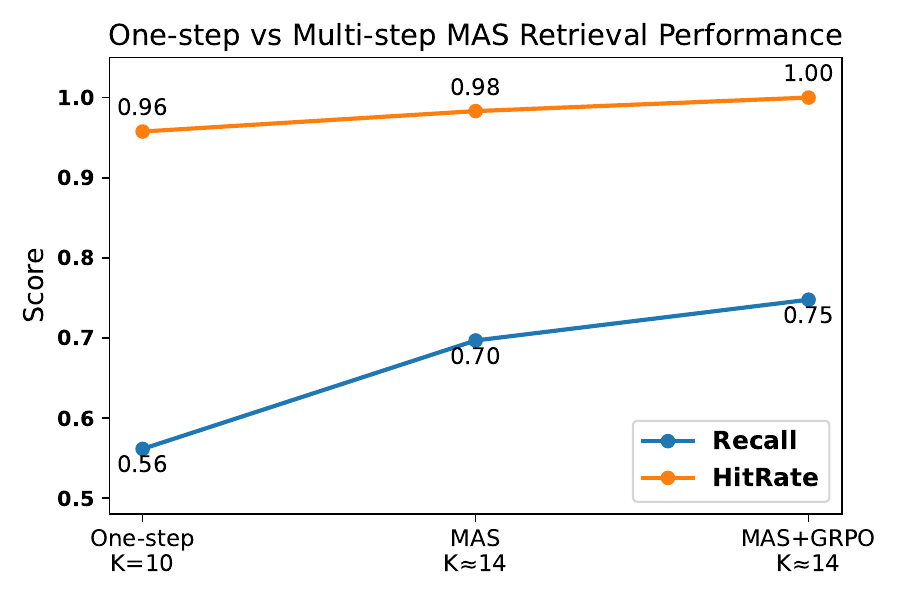}
    \caption{
    Ablation of iterative search depth in MAS. The One-step configuration executes only a single rewrite and retrieves a fixed top--10 list, whereas multi-step MAS and MAS+GRPO operate with dynamic candidate pool sizes (approximately 14 on average). Multi-step search substantially improves coverage, and GRPO further enhances retrieval effectiveness.
    }
    \label{fig:mas_one_step}
\end{figure}

We further examine the contribution of iterative search by comparing three retrieval configurations: a single-step MAS (One-step), a multi-step MAS without GRPO training, and the full MAS with GRPO policy. As shown in Figure~\ref{fig:mas_one_step}, One-step MAS achieves a Recall@10 of 0.56 and a HitRate@10 of 0.96. This behaviour aligns with the observations in Appendix~\ref{app:case_study}, where the first rewrite produced by MAS often acts as a directional refinement of the user query. Such refinements sharpen the semantic focus of the query and help surface some relevant statutes, but they do not expand the coverage sufficiently to recover all relevant provisions. When the retrieval process is truncated after this first refinement, MAS lacks the opportunity to issue complementary reformulations that would compensate for missing semantic aspects of the query, resulting in lower recall.

Allowing MAS to perform multiple iterations immediately improves retrieval coverage. The untrained multi-step MAS achieves a Recall@14 of 0.70 and a HitRate@14 of 0.98, indicating that later rewrites naturally address coverage gaps introduced by the initial directional rewrite. Although this configuration requires more computation on complex queries, MAS terminates early on simpler ones. The average embedding budget therefore remains close to the single-step setting, consistent with the compute analysis in Section~\ref{sec:exper_setup}. This adaptive allocation of retrieval depth is a fundamental advantage of the MAS design.

After GRPO optimisation, MAS with GRPO further increases Recall@14 to 0.75 and achieves a perfect HitRate@14. These improvements are consistent with the intended effect of GRPO, which encourages the policy to allocate retrieval depth more effectively and stabilise rewrite selection and termination decisions. The learned policy tends to employ additional iterations when the initial rewrite does not yield sufficient coverage, while stopping earlier on queries that are already well resolved. This leads to a more complete and better structured candidate pool.

Together, these results show that the effectiveness of MAS relies on its multi-step structure. The largest gains emerge when multiple semantically diverse reformulations are aggregated. One-step MAS provides only a partial view of the legal interpretation space, whereas multi-step MAS reconstructs complementary perspectives that substantially improve recall. As illustrated by the representative trajectories in Appendix~\ref{app:case_study}, MAS with GRPO both recovers statutes missed by the initial rewrite and terminates efficiently on simpler queries. These findings confirm that iterative search is essential for achieving the high-recall behaviour required for robust statute retrieval.

\subsection{Cost and Efficiency Analysis}

Although LegalMALR introduces an iterative retrieval mechanism, its overall computational profile remains comparable to that of conventional RAG pipelines. Empirically, MAS triggers an average of 2.01 dense-retrieval calls per query during evaluation on the STARD and CSAID datasets, with a maximum of four retrieval calls observed in any trajectory. Retrieval calls are not equivalent to planner iterations: a single planner step may issue multiple retrievals through the Multi-Element Decomposition Agent or may issue none when producing an exit decision. In practice, however, no evaluation trajectory exceeded four retrieval calls. Each retrieval call returns the top thirty embedding candidates, which are pruned to the top ten using the lightweight reranker. This yields an average of approximately sixty embedding retrieval operations per query, a value that closely matches the standard RAG configuration that retrieves a fixed top sixty candidates once per query.

The reranking cost is also well bounded. Lightweight reranking is applied only within MAS to prune intermediate candidate lists. The final large model based reranking stage processes an average of 13.58 merged and deduplicated candidate statutes rather than the full embedding set. Although the LLM reranker in LegalMALR is more capable than the lightweight rerankers used in baseline RAG systems, the substantial reduction in input size limits its token consumption and offsets the higher per-token cost of the larger model. As a result, the end-to-end reranking overhead remains practical and does not dominate total inference time.

Overall latency increases moderately on complex queries due to iterative reformulation, but MAS terminates early on simpler queries, resulting in an average retrieval depth that remains close to the single-step setting. This adaptive allocation of computation is a key advantage of the MAS design: more retrieval effort is invested only when necessary, while redundant expansions are suppressed. GRPO optimisation further stabilises this behaviour by reducing unnecessary iterations and encouraging more consistent retrieval trajectories. These observations show that LegalMALR achieves substantial gains in retrieval quality while maintaining a competitive and resource-efficient computational profile.

\section{Conclusion and Future Work}\label{sec:conclusion}

Legal statute retrieval presents unique challenges due to the implicitness, multi issue structure, and doctrinal complexity of real world legal queries. Many queries describe factual scenarios rather than citing statutory terminology, and often reference multiple statutory elements that are only partially or implicitly expressed in natural language. These characteristics make accurate retrieval substantially more difficult than conventional single issue or keyword based search and require retrieval systems capable of reconstructing latent legal structure while ensuring high quality ranking among candidate statutes.

\subsubsection{Summary of the Proposed Framework}

This work introduces LegalMALR, a retrieval framework that combines a multi agent query understanding system with a large language model based reranking module. The MAS generates diverse and complementary reformulations of the input query and supports iterative retrieval through a planner controlled decision process. To stabilise the inherently stochastic behaviour of LLM based reformulations, we train a unified MAS policy using Generalized Reinforcement Policy Optimization, which encourages trajectories that reveal statutory elements relevant to retrieval. The merged candidate set produced by the MAS is then evaluated by a commercial large language model, Qwen-Max, which performs natural language legal reasoning to assess statutory applicability. Together, these components form a coherent system that jointly enhances recall, interpretive robustness, and final ranking accuracy.

\subsubsection{Summary of Empirical Findings}

Our experiments on both STARD and CSAID demonstrate that the proposed framework substantially improves statute retrieval performance under both in distribution and out of distribution settings. The MAS yields broad candidate coverage by exploring multiple doctrinal interpretations of the query, and GRPO training significantly reduces rollout variability while improving the completeness of the retrieved candidate set. The LLM reranker provides strong gains in ranking quality, consistently outperforming lightweight supervised rerankers across metrics such as MRR, nDCG, and Recall at K. These results confirm that combining multi perspective query understanding, reinforcement based policy optimisation, and large model reranking yields clear benefits over conventional RAG pipelines and dense retrieval baselines.

\subsubsection{Limitations}

Despite its strong performance, the framework has several limitations. First, the multi agent query understanding process requires multiple LLM calls and may increase computational cost for complex queries, although GRPO training effectively constrains the average retrieval depth to a modest level. Second, the system relies on an embedding based retriever for initial candidate collection, and its performance may depend on the quality of underlying vector representations. Third, although the Qwen-Max reranker provides stronger statutory reasoning than lightweight models, it remains an inference time component without task specific tuning and may not capture all facets of doctrinal nuance. Finally, the datasets used in this study focus on Chinese statutes and Chinese language queries, and further work is needed to assess how well the approach generalises to other jurisdictions and legal traditions.

\section{Discussion}\label{sec:discussion}

This section summarises the practical considerations of the proposed framework and outlines both its current limitations and the directions that we believe are most promising for future research. While LegalMALR substantially improves recall and ranking quality under both in-distribution and cross-domain settings, several technical and methodological constraints remain. Addressing these constraints will help further strengthen the robustness, efficiency, and applicability of multi-agent legal retrieval systems.

\subsection{Limitations}\label{sec:limitations}

Although LegalMALR demonstrates strong performance across multiple evaluation settings, several limitations warrant attention.

First, the behaviour of MAS remains sensitive to large language model variability. The rewrite agents occasionally generate incomplete or overly specific reformulations, and the decomposition agent may either miss opportunities for structural decomposition or introduce unnecessary fragments. GRPO improves stability but cannot eliminate stochasticity entirely because the reward signal is sparse and tied only to trajectory-level recall.

Second, the reliance on a powerful commercial LLM for the final reranking stage introduces additional computational cost and limits full reproducibility. While the candidate set processed by the LLM is relatively small, the per-token cost is still higher than that of lightweight rerankers. In addition, the proprietary nature of commercial LLMs restricts fine-grained control over reasoning behaviour and complicates long-term model maintenance.

Third, multi-step retrieval unavoidably increases latency in complex queries. MAS mitigates this through adaptive termination, yet the worst-case number of retrieval calls is still higher than that of single-step pipelines. In real-time or high-throughput applications, this may require batching, caching, or further optimisation.

Fourth, although the STARD and CSAID datasets capture diverse legal search scenarios, they do not represent the full breadth of jurisdictions, procedural contexts, or specialised domains in real-world legal practice. The relatively small size of CSAID also limits the statistical robustness of certain analyses. Broader evaluations are needed to understand the generalisability of multi-agent retrieval systems in more varied legal settings.

A further limitation is that the entire framework is developed and evaluated within a Chinese language setting. All training data, statutory corpora, MAS reformulations, and LLM reranking prompts operate exclusively on Chinese legal texts. The effectiveness of MAS-driven reformulation and GRPO optimisation therefore reflects linguistic and structural characteristics specific to Chinese statutory language. The cross-lingual robustness of LegalMALR remains unexplored, and it is unclear to what extent the current multi-agent design transfers to jurisdictions that differ in drafting conventions, logical structure, or linguistic expression. Extending the framework to multilingual or cross-jurisdictional retrieval constitutes an important direction for future research.

Finally, the reward formulation used in GRPO focuses exclusively on MAS-stage recall. This metric is aligned with our objective of improving coverage, but it does not explicitly model ranking quality, interpretability of rewrite decisions, or the legal validity of intermediate reasoning steps. Future optimisation frameworks could benefit from richer feedback signals.

These limitations do not undermine the empirical findings of this work but instead highlight the opportunities for further refinement and extension of the LegalMALR framework.

\subsection{Future Directions}\label{sec:future_directions}

Several promising research directions emerge from this work. One direction concerns data distillation for legal retrieval. High-quality statutory annotations are costly to produce and often exhibit inconsistency. The multi-step reformulations, retrieval trajectories, and final rankings produced by LegalMALR offer a rich source of supervision for training smaller embedding models and lightweight rerankers. Distilling multi-step legal reasoning into compact retriever architectures may enable systems that approximate the performance of LegalMALR while achieving substantially lower inference cost.

A second direction involves simplifying and accelerating the multi-agent retrieval pipeline. Although iterative reformulation contributes significantly to recall gains, it also increases latency for complex queries. Reducing the number of rewrite and retrieval steps without degrading performance remains an important objective. Potential approaches include learning compressed one-step approximations of MAS behaviour, improving the planner’s ability to predict when additional reformulations are unnecessary, or replacing multi-step search with a single jointly optimised retrieval policy.

A third direction explores the integration of LegalMALR as a retrieval component within broader agentic AI systems. In long-horizon tasks such as due diligence, compliance auditing, and multi-stage legal research, retrieval accuracy and doctrinal grounding are more important than per-step latency. The interpretive flexibility of MAS combined with the deeper legal reasoning capability of the LLM reranker makes LegalMALR a natural candidate for incorporation into multi-agent reasoning frameworks, particularly those requiring grounded statutory understanding across extended workflows.

Together, these directions highlight the broader potential of LegalMALR as a foundation for future research at the intersection of statutory retrieval, multi-agent optimisation, and large-model-based legal reasoning.

\section{Conclusion}

This work introduces LegalMALR, a multi-agent legal retrieval framework that combines iterative query reformulation, GRPO-optimised policy learning, and large model based reranking. The system addresses the challenges posed by implicit, multi element, and colloquial legal queries, and achieves substantial gains over strong dense retrieval and RAG baselines on both in-distribution and cross-domain evaluations. The MAS module expands recall through structured reformulations, while GRPO stabilises its behaviour and improves coverage. The final LLM reranker provides principled statutory relevance assessment and consistently enhances ranking quality. Together, these components enable LegalMALR to deliver reliable, high recall statute retrieval under realistic legal search conditions. We hope that this framework will serve as a foundation for future advances in multi-agent optimisation and legally grounded retrieval systems.

\section*{Statements and Declarations}

\subsection*{Competing Interests}
The authors have no relevant financial or non-financial interests to disclose.

\subsection*{Funding}
No funding was received for conducting this study.

\subsection*{Data Availability}
The CSAID dataset is publicly available at \url{https://github.com/lyxx3rd/CSAID}.

\subsection*{Code Availability}
The code supporting the findings of this study is publicly available at \url{https://github.com/lyxx3rd/LegalMALR}.

\bibliography{sn-bibliography}

\clearpage
\appendix
\onecolumn

\section{Representative Examples from CSAID}\label{app:CSAID}

This appendix provides two representative examples from the CSAID dataset. Both instances demonstrate typical challenges in real-world Chinese legal search, including implicit factual conditions, domain-specific jargon, and doctrinal structures that diverge from surface-level semantics. These examples illustrate why conventional semantic retrieval may fail and why multi-perspective query interpretation is essential. These examples are presented in Table~\ref{tab:csaid_examples}.

\begin{table*}[htb!]
\centering
\small
\begin{tabularx}{\textwidth}{p{1.8cm} X}
\toprule
\textbf{Example A} &
\textbf{User Query:} \textit{Will I be criminally punished if I ate a protected animal under conditions of extreme hunger?}

\vspace{1mm}
\textbf{Surface Meaning:}  
Appears to concern wildlife-protection or food-safety legislation.

\textbf{Underlying Legal Issue:}  
The true doctrinal focus is the \textit{defence of necessity} in Chinese criminal law.  
Relevant statutory applicability depends on implicit conditions not stated in the query, including:
\begin{itemize}
    \item the existence of imminent danger,
    \item proportionality between harm avoided and harm caused,
    \item whether alternative options existed.
\end{itemize}

\textbf{Significance:}  
The gap between literal wording and doctrinal inference demonstrates why similarity-based retrieval often fails for such queries. \\
\midrule

\textbf{Example B} &
\textbf{User Query:} \textit{How is the legal validity of the commonly used “back-to-back clause” in engineering project contracts determined?}

\vspace{1mm}
\textbf{Surface Meaning:}  
Uses industry-specific jargon (“back-to-back clause”) that does not appear in statutory language.

\textbf{Underlying Legal Issue:}  
The actual concern is the validity of \textit{conditional payment obligations} in contract law, where payment duties depend on third-party actions—an interpretation not recoverable via direct semantic matching.

\textbf{Significance:}  
This example illustrates how domain-specific terminology conceals doctrinal meaning, causing conventional retrievers to miss the correct legal provisions. \\
\bottomrule
\end{tabularx}
\caption{Representative CSAID examples illustrating implicit factual conditions and domain-specific conceptual structures.}
\label{tab:csaid_examples}
\end{table*}

\clearpage
\section{MAS Agent System Prompts}
\label{app:agent_prompts}

This appendix provides the system level specifications that instantiate the agent roles defined in the MAS framework. While the conceptual functions of the agents are described in the main text, the operational behaviour of each agent is governed by a system prompt that assigns a distinct legal interpretation perspective and determines how the agent contributes to the multi step retrieval process. Table~\ref{tab:agent_prompts} summarises the functional content of these system prompts.

All system prompts originate from Chinese instructions used during inference. For reproducibility, the table presents faithful English summaries that preserve the functional intent of the original prompts while omitting formatting patterns and implementation specific phrasing that do not influence model behaviour. Each summary captures the core decision rule, the reformulation strategy, the legal perspective applied by the agent and the expected output structure. These summaries allow researchers to reconstruct the operational behaviour of the MAS module without requiring verbatim disclosure of long form instructional text.

The system prompts form the only source of differentiation among agents, since all agents share the same Qwen 3 4B Instruct backbone. They specify the reformulation logic used by the rewrite agents, the analytical criteria applied by the abnormality analysis agents and the control policy executed by the planner. When used together, these prompts enable the MAS to explore complementary factual and doctrinal interpretations of the query, enhance retrieval coverage and support reliable multi round decision making. The summaries in Table~\ref{tab:agent_prompts} therefore provide a complete and self contained description of the operational constraints that define MAS agent behaviour during inference.

\begin{table*}[t]
\centering
\caption{Functional summaries of the system prompts used by MAS agents.}
\label{tab:agent_prompts}
\begin{tabularx}{\textwidth}{p{2.8cm} X}
\toprule
\textbf{Agent} & \textbf{System Prompt Summary (English)} \\
\midrule

Planner Agent & Analyse the query and previously retrieved statutes. Select the next action from a fixed set of rewrite or analysis agents. Encourage multi round retrieval, identify missing legal elements, determine when new reformulations are needed and decide when to terminate the search. Return an action choice and a concise justification. \\
\midrule
Single Element Rewrite Agent & Rewrite the query into precise legal language while preserving meaning. Clarify colloquial expressions and improve legal term specificity. Produce one refined reformulation suitable for statute retrieval. \\
\midrule
Supplementary Element Rewrite Agent & Add implicit but legally decisive factual or normative conditions to the query. Make thresholds, actor identities, obligations or contextual qualifiers explicit to address missing statutory applicability elements. Produce one targeted reformulation. \\
\midrule
Multi Element Decomposition Agent & Identify queries containing multiple legal elements, subjects or doctrinal components. Decompose the query into several focused sub queries that can be retrieved independently. Output multiple reformulations when necessary, each addressing a distinct legal element. \\
\midrule
Supportive Law Rewrite Agent & Generate reformulations that retrieve interpretive, procedural or auxiliary statutes that offer contextual or doctrinal support for the core substantive provisions. Maintain the original meaning while expanding the legal reasoning context. \\
\midrule
Semantic Abnormality Analyzer & Detect semantic anomalies in the query such as vitiation of intent, domain overlap, latent principles or procedural dependencies. Classify the anomaly type and provide a brief explanation. Output a structured JSON diagnosis. \\
\midrule
Semantic Abnormality Rewriter & Reformulate queries that contain semantic anomalies into legally coherent expressions suitable for retrieval. Retain the original legal meaning while clarifying domain, logic or doctrinal direction. Output a repaired query and a short explanation. \\

\bottomrule
\end{tabularx}
\end{table*}

\clearpage
\section{LLM Reranker Prompt Specification}
\label{app:reranker_prompt}

The LLM reranker evaluates the merged candidate statutes produced by MAS and produces a final relevance ranking based on substantive legal reasoning. The reranker receives both the user query and a list of candidate statutes, each accompanied by a model generated relevance score. It must consider but not mechanically follow these scores and must perform an independent legal analysis of statutory applicability. The reranker is implemented with Qwen-Max, which is instructed through a system prompt that defines its evaluative role, and a user prompt template that provides the query, the candidate statutes and the expected output format. Functional summaries of these prompts are provided in Table~\ref{tab:reranker_prompts}. These summaries preserve the operational intent of the original prompts while omitting implementation specific phrasing.

\begin{table*}[h]
\centering
\caption{Functional summaries of the system and user prompts used by the LLM reranker.}
\label{tab:reranker_prompts}
\begin{tabularx}{\textwidth}{p{3.2cm} X}
\toprule
\textbf{Prompt Type} & \textbf{Functional Summary (English)} \\
\midrule

System Prompt & The model is assigned the role of a legal expert who must evaluate candidate statutes using both objective relevance scores and substantive legal reasoning. It must weigh statutory applicability, assess factual alignment with the query and avoid relying solely on numerical scores. It must provide a ranking that reflects legal sufficiency, doctrinal relevance and completeness of the solution. The model is instructed to output results in a strict JSON format containing the indices of the selected statutes. \\

\midrule
User Prompt Template & Provides the user query, the complete list of retrieved candidate statutes and their reranker scores. Instructs the model to follow a multi stage evaluation procedure: interpret score meaning, perform legal analysis of each statute, balance score based and law based relevance, select the top \(k\) statutes and verify coverage. The output must be a JSON dictionary containing the selected indices. The template does not influence decision logic beyond providing task context and required output format. \\

\bottomrule
\end{tabularx}
\end{table*}

\clearpage
\section{Case Studies of MAS+GRPO Retrieval Trajectories}
\label{app:case_study}

\begin{figure*}[h]
    \centering
    \includegraphics[width=0.8\linewidth]{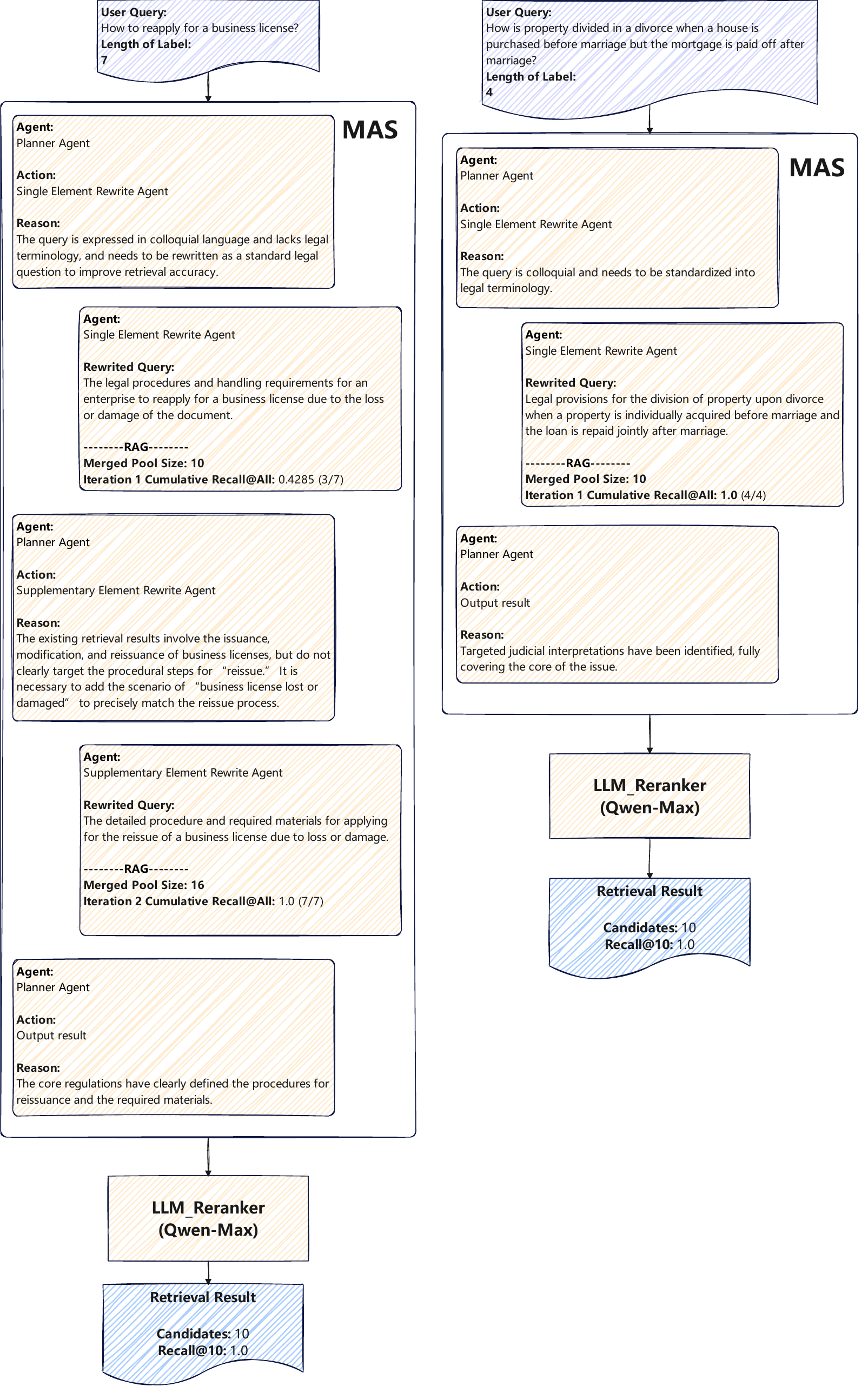}
    \caption{Combined case study of MAS+GRPO behaviour. The left example illustrates a multi step retrieval trajectory in which the second reformulation exposes complementary legal elements and achieves full statutory coverage. The right example illustrates early stopping behaviour where the initial reformulation already captures the necessary statutory elements and the system terminates the iterative process. Recall values appear only as diagnostic indicators.}
    \label{fig:case_combined}
\end{figure*}

This appendix presents two representative retrieval trajectories produced by the MAS+GRPO framework. These trajectories illustrate how the system adapts its iterative behaviour according to query complexity. The left example demonstrates a multi step process in which later reformulations recover statutes missed by the initial rewrite. The right example shows early termination on a simpler query in which the first reformulation already provides complete statutory coverage. These retrieval trajectories are illustrated in Figure~\ref{fig:case_combined}.

The recall values shown in the figure are diagnostic signals used solely to illustrate coverage changes across iterations. MAS does not have access to recall during inference and does not rely on recall information for decision making. The Reason fields in the illustration represent natural language explanations generated by MAS agents as part of their decision outputs. For readability these explanations have been lightly condensed while preserving their substantive intent.

\end{document}